\documentclass[pdflatex,sn-basic,Numbered]{sn-jnl}

\usepackage{amsmath,amssymb,amsfonts}
\usepackage{graphicx}
\usepackage{textcomp}
\usepackage{xcolor}
\usepackage{booktabs}
\usepackage{multirow}
\usepackage{array}
\usepackage{listings}
\usepackage{enumitem}
\usepackage{longtable}
\usepackage{hyperref}
\usepackage{tikz}
\usetikzlibrary{positioning, arrows.meta}

\definecolor{mygray}{rgb}{0.5,0.5,0.5}
\definecolor{myblue}{rgb}{0.1,0.1,0.9}
\definecolor{mylightgray}{rgb}{0.96,0.96,0.96}

\begin{document}

\title[Reference-free retrieval tuning for ESG KPI extraction]{A Multi-Layer Reference-Free Completeness Framework for Retrieval Tuning in ESG KPI Extraction}

\author[1]{\fnm{Kianna} \sur{Kazemi}}\email{kianna.kazemi@ifabfoundation.org}
\author*[2]{\fnm{Motaz} \sur{Saad}}\email{motazk.saad@unisalento.it}
\author[1]{\fnm{Ivan} \sur{Gentile}}\email{ivan.gentile@ifabfoundation.org}
\author[2]{\fnm{Antonella} \sur{Longo}}\email{antonella.longo@unisalento.it}

\affil[1]{\orgname{IFAB Foundation}, \orgaddress{\city{Bologna}, \country{Italy}}}
\affil*[2]{\orgname{University of Salento}, \orgaddress{\city{Lecce}, \country{Italy}}}

\abstract{
Retrieval-augmented generation (RAG) systems for ESG KPI extraction depend strongly on retrieval settings such as chunk size and retrieval depth, but selecting these settings is difficult when ground-truth answers are unavailable at scale. This paper studies retrieval tuning in an extractive ESG RAG pipeline using a reference-free evaluation framework based on data completeness. We define a multi-layer completeness framework that combines response-level completeness, field-level completeness for structured outputs, disclosure-conditional completeness, and a refusal taxonomy. We apply this framework to a parameter sweep over chunk size and top-k retrieval on a stratified set of European sustainability reports and ESG questions. The results show that retrieval parameters materially affect completeness and cost, and that the proposed framework provides a practical way to compare configurations without annotated reference answers. The framework measures output completeness rather than factual accuracy. A post-hoc, evidence-grounded multi-model audit diagnoses failure modes on a bounded sample, but it is not a reference-answer accuracy benchmark. The study supports the use of multi-layer reference-free completeness as a configuration-selection signal for ESG KPI extraction pipelines.
}

\keywords{retrieval-augmented generation, ESG, sustainability reports, KPI extraction, reference-free evaluation, retrieval tuning}

\maketitle

\section{Introduction}
\label{sec:intro}

The Corporate Sustainability Reporting Directive (CSRD) and the convergence of global ESG-disclosure frameworks (GRI, SASB, TCFD) have turned corporate sustainability reports into high-volume, regulator-facing documents whose structured KPIs---scope-1/2/3 emissions, workforce composition, governance metrics, audited financials---feed downstream reporting databases. Retrieval-augmented generation (RAG)~\cite{lewis2020rag} with open-source LLMs has emerged as the dominant approach: reports are converted from PDF to Markdown, indexed in a vector store, and queried by an LLM grounded on the retrieved chunks.

In production, the cost of \emph{choosing} a RAG configuration competes with the cost of \emph{running} it. Reference-based evaluation frameworks---RAGAS~\cite{es2024ragas}, RAGChecker~\cite{ru2024ragchecker}, BERGEN~\cite{rau2024bergen}---require either large LLM judges or human-annotated reference answers. For sustainability reports, the latter is prohibitive: the canonical question set spans 59 GRI-aligned KPIs across narrative, tabular, and mixed-content disclosures, and creating a human-annotated reference set for a corpus of 14{,}389 documents would take months. Practitioners therefore default to one of two shortcuts: (i) port a configuration from an unrelated domain~\cite{rau2024bergen}, or (ii) inspect a small sample manually. Neither approach generalizes.

This paper proposes a reference-free, multi-layer completeness framework for comparing retrieval configurations in ESG KPI extraction and uses it to study the effect of chunk size and top-$k$ on extraction quality. We instantiate the framework through four layers of completeness (Layer~1 strict null with response-shape check, Layer~2 field-level JSON fill-rate, Layer~3 disclosure-conditional recall via cross-configuration agreement, Layer~4 refusal taxonomy) and evaluate it on a $4 \times 4$ grid of chunk-size and retrieval-depth values on a 300-document stratified sustainability subset under a fixed open-source generator (Qwen3-8B~\cite{qwen3report}) and a fixed retriever (Weaviate~\cite{weaviate2024}, HNSW).

\subsection*{Contributions}
\begin{enumerate}
  \item \textbf{A multi-layer reference-free completeness metric (C1).} We define four layers that progressively refine the binary ``answer / no answer'' question into: response-level strict null, field-level JSON fill-rate, disclosure-conditional recall against a cross-configuration majority-vote proxy for ``actually disclosed'', and a refusal taxonomy. Each layer is computable from the raw extraction CSV without invoking any model at evaluation time.

  \item \textbf{A retrieval-axis sweep methodology for ESG extraction at scale (C2).} We isolate the retrieval-parameter axis of a production-grade RAG pipeline---holding generator, embedding model, prompt template, and corpus fixed---and characterize its 16-cell behavior on a stratified 300-document subset spanning seven sectors, fourteen EU jurisdictions, and a 29-KPI catalog drawn proportionally from the GRI master.

  \item \textbf{A Pareto frontier in (multi-layer completeness, context cost, latency) (C3).} We chart the frontier under two cost-vs.-quality axes (L1 vs.\ context tokens and L3 vs.\ context tokens) and identify the Pareto-dominated cells. On the full 16-cell grid, ten cells are non-dominated under L1 and ten under L3, and the steepest gradient occurs between chunk~256 / top-$k$~10 and chunk~256 / top-$k$~40.

  \item \textbf{A refusal taxonomy that reveals a previously-hidden failure mode (C4).} The Layer-4 classifier shows that \emph{shape mismatches} account for 6--10\,\% of all responses across the grid: these are responses that do not satisfy the numeric or yes/no shape contract of the question. This category is invisible to literal-NA detectors and accounts for roughly one-fifth to one-quarter of the L1 gap.
\end{enumerate}

\subsection*{Positioning within the project}
This paper is one of three studies of a production ESG KPI-extraction pipeline built from public sustainability disclosures. The preprint by Saad et al.~\cite{companion_a_saad2026} holds retrieval fixed and evaluates seven open-source generators on a 498-report corpus with RAGAS, finding substantial variation in generation-side metrics while context recall remains well below unity under top-1 retrieval. The Preprints.org preprint by Gentile et al.~\cite{companion_b_gentile2026} holds the generator class fixed and varies the \emph{corpus representation}: a vision--language augmentation stage appends structured descriptions of chart and infographic content that Docling-style text conversion discards, increasing recoverable numeric tokens in the indexed Markdown before any RAG query is run. The present paper holds both the generator (Qwen3-8B) and the text-only Docling corpus fixed and isolates the \emph{retrieval-configuration} axis---chunk size and top-$k$---using a reference-free completeness metric rather than RAGAS. The three studies address complementary axes (generator choice, corpus modality, and retrieval tuning): VL augmentation can enrich the corpus on which any retrieval grid is run, while generator benchmarking and retrieval sweeps answer different configuration questions for the resulting text index. All three share the Qwen3-Embedding-4B dense retriever. The preprint by Saad et al. evaluates fourteen embedding models for ESG retrieval and RAG.

\section{Related Work}
\label{sec:related}

This section is a focused narrative synthesis rather than a systematic literature review; it does not claim exhaustive coverage. The cited works were selected for their direct relevance to RAG evaluation, abstention, and ESG-information extraction.

\paragraph{RAG and its hyperparameters.}
Lewis et al.~\cite{lewis2020rag} introduced the canonical RAG pattern. Since then, retrievers (BM25, dense bi-encoders, learned sparse, hybrid), rerankers (cross-encoders, LLM rerankers), and chunking strategies (fixed-size, semantic, structure-aware) have each spawned their own literature. BERGEN~\cite{rau2024bergen} consolidates 500+ open-domain QA experiments on Wikipedia-style collections and finds that retrieval quality and reranking strongly affect downstream generation, while noting inconsistent benchmarks and limited coverage of domain-specific corpora. In contrast, on long, semi-structured, regulator-facing sustainability reports, with the generator held fixed, we show that retrieval-configuration parameters alone---chunk size and top-$k$---move strict response-level completeness by 24~pp and disclosure-conditional completeness by 27~pp across all 16 cells (\S\ref{sec:results}).

\paragraph{Reference-free RAG evaluation.}
RAGAS~\cite{es2024ragas} automates RAG assessment with metrics such as faithfulness and answer relevance (LLM-based claim checking and embedding similarity, respectively) and, in later releases, context precision and recall (typically LLM-judged against a reference answer). The framework reduces reliance on human-annotated gold answers at evaluation time, but each score still incurs LLM and/or embedding API cost. RAGChecker~\cite{ru2024ragchecker} refines the same RAG-triad idea to \emph{claim-level} entailment between a response, its retrieved context, and a ground-truth answer, yielding finer diagnostics but still requiring an LLM checker and reference answers for several metrics. Self-RAG~\cite{asai2024selfrag} is orthogonal: it trains the \emph{generator} to emit reflection and critique tokens at inference, rather than supplying an offline evaluator. Our setting instead uses rule- and pattern-based detectors over the raw extraction CSV---no LLM, embeddings, or human labels at scoring time. \emph{Data completeness} is that family of metrics (\S\ref{sec:completeness}); the trade-off is the set of blind spots documented in \S\ref{sec:limits}.

\paragraph{Abstention and refusal in QA.}
A growing line of work studies when language models should \emph{abstain} rather than answer---through selective prediction, calibrated rejection thresholds, and RAG systems that learn to withhold unsupported generations. Self-RAG~\cite{asai2024selfrag} trains the generator to emit critique tokens that score passage relevance and whether an output is supported by retrieved evidence, enabling adaptive retrieval and self-filtering at inference. In high-stakes document QA, explicit refusals are often preferable to confident errors: FinanceBench~\cite{islam2023financebench} finds that vector-augmented models frequently fail by answering incorrectly rather than by declining to answer, so ``failure to answer'' is a first-class outcome worth measuring alongside accuracy. We track the complementary quantity: \emph{data completeness} is the share of (document, question) pairs that receive a non-null extraction under our detectors (\S\ref{sec:completeness}). Layer~1 approximates a strict \emph{answer rate}---refusals such as ``I don't know'' or ``not specified in the provided context'' count as null---but also flags literal-NA responses, infrastructure errors, and shape-contract violations (e.g.\ prose where a number was required). Optimizing completeness alone cannot distinguish a correct extraction from a confident hallucination; that limitation is central to \S\ref{sec:limits}.

\paragraph{ESG / sustainability extraction.}
ESGReveal~\cite{zou2023esgreveal} reports 76.9\,\% data-extraction accuracy for GPT-4 on a stratified sample of 166 Hong Kong Exchange--listed companies (2022), using proprietary models, structured ESG metadata, and RAG over HK disclosures. On EU-Taxonomy-compliant KPI extraction, \cite{schmoll2025eutaxonomy} find in a zero-shot setting a \emph{paradox of context}: concise structured metadata can outperform feeding the full unstructured annual report for activity classification, even though models fail on quantitative KPI regression without supervision. Our sweep probes a different lever---how much report text to retrieve (chunk size and top-$k$)---and on the completed grid completeness is generally non-decreasing in both axes (\S\ref{sec:monotonicity}), so more retrieved context usually yields more non-null answers rather than less. FinanceBench~\cite{islam2023financebench} provides a finance-domain stress test: on its 150-question human-eval subset, GPT-4-Turbo with a \emph{shared} document vector store is correct on only 19\,\% of items (81\,\% incorrect or explicit refusal in the paper's headline figure), reminding us that a high non-null rate does not imply accuracy and that whether the model answers at all remains a basic diagnostic.

\section{System under Study}
\label{sec:system}

\begin{figure}[t]
\centering
\begin{tikzpicture}[
  node distance=0.45cm and 0.5cm,
  box/.style={draw=blue!60, fill=blue!10, thick, rounded corners=2pt, minimum height=0.55cm, minimum width=2.2cm, text centered, font=\scriptsize},
  data/.style={draw=orange!70, fill=orange!10, thick, minimum height=0.5cm, minimum width=1.9cm, text centered, font=\scriptsize},
  varied/.style={draw=red!70, fill=red!10, thick, minimum height=0.5cm, minimum width=1.9cm, text centered, font=\scriptsize, dashed},
  arrow/.style={thick, ->, >=Stealth}
]
\node[data] (pdf) {PDF reports};
\node[box, right=of pdf] (docling) {Docling};
\node[data, right=of docling] (md) {Markdown};
\node[varied, below=of md] (chunker) {Token chunker};
\node[box, below=of chunker] (embed) {Qwen3-Embedding-4B (vLLM)};
\node[data, left=of embed] (weaviate) {Weaviate / HNSW};
\node[varied, left=of weaviate] (retr) {top-$k$ retriever};
\node[box, below=of weaviate] (qwen) {Qwen3-8B (vLLM, 1 worker)};
\node[data, right=of qwen] (resp) {LLM response};
\draw[arrow] (pdf) -- (docling);
\draw[arrow] (docling) -- (md);
\draw[arrow] (md) -- (chunker);
\draw[arrow] (chunker) -- (embed);
\draw[arrow] (embed) -- (weaviate);
\draw[arrow] (weaviate) -- (retr);
\draw[arrow] (retr) -- (qwen);
\draw[arrow] (qwen) -- (resp);
\end{tikzpicture}
\caption{Pipeline under study. Dashed boxes indicate components varied in the sweep (chunk size in the chunker, top-$k$ in the retriever). All other components are fixed.}
\label{fig:pipeline}
\end{figure}
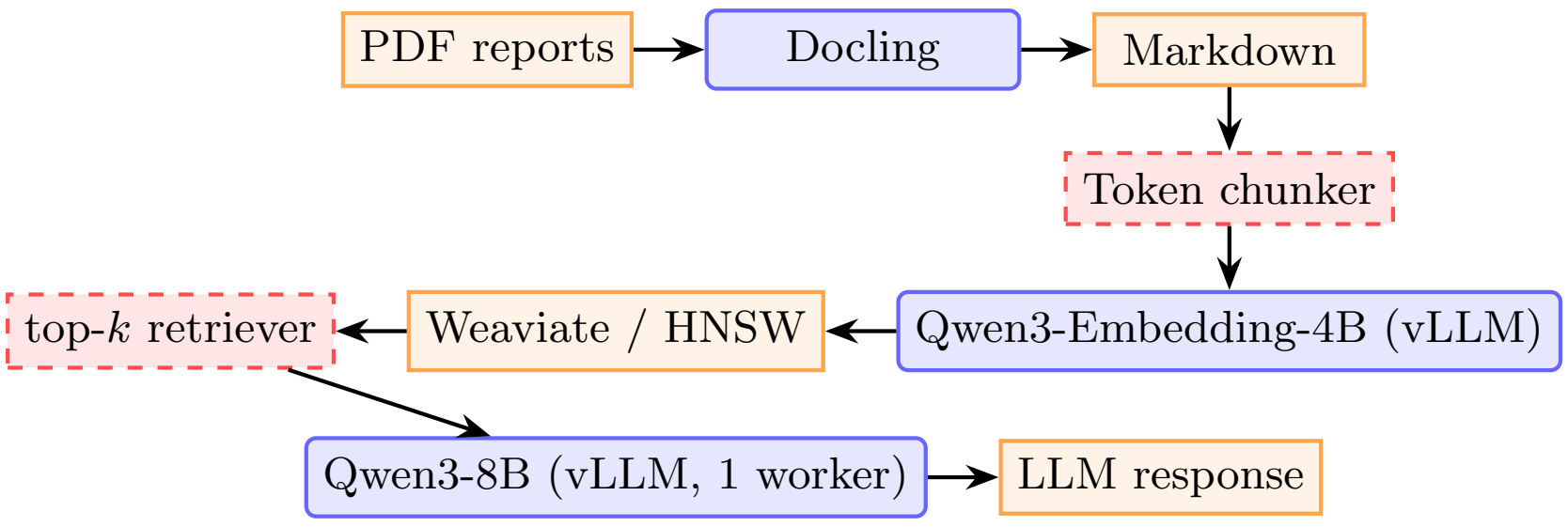

The pipeline has two phases (Figure~\ref{fig:pipeline}): an indexing phase that runs once for each chunk size and a query phase that reuses the persistent index. All swept configurations share the components below; only chunk size and top-$k$ vary.

\paragraph{Document ingest.}
PDFs are converted to GitHub Flavored Markdown by Docling~\cite{docling2024} with table extraction in \texttt{ACCURATE} mode and RapidOCR fallback for scanned pages. The stratified subset comprises 300 documents and produces 300 Markdown files, which LlamaIndex's \texttt{SimpleDirectoryReader} loads.

\paragraph{Chunking.}
Documents are split with LlamaIndex's \texttt{TokenTextSplitter}. Chunk overlap co-varies with chunk size at a constant $\approx 15\%$ ratio, yielding $\{38, 77, 154, 307\}$ tokens for chunk sizes $\{256, 512, 1024, 2048\}$. One chunk-filter parameter is held fixed throughout: a minimum chunk length of 100 characters (chunks shorter than this are dropped). Holding overlap proportional rather than factorial keeps re-indexing tractable (one fresh index per chunk size; query-time parameters reuse the existing index).

\paragraph{Embedding and index.}
We fix the embedder to Qwen3-Embedding-4B. The preprint by Saad et al.~\cite{saad2026esgembed} evaluates fourteen embedding models for ESG retrieval and RAG, with Qwen3-based models achieving the strongest overall performance; the present study does not independently re-evaluate that selection. Chunks are embedded with that model served by vLLM~\cite{kwon2023vllm} (bfloat16, \texttt{gpu\_memory\_utilization}\,=\,0.50, dedicated GPU~0) and indexed in an embedded Weaviate~\cite{weaviate2024} v4 collection with an HNSW vector index (cosine distance, \texttt{ef\_construction}\,=\,128, \texttt{ef}\,=\,$-1$ (dynamic), \texttt{maxConnections}\,=\,32---all Weaviate defaults). Embedding requests are batched in groups of 32; chunks are committed to Weaviate in batches of 50; eight CPU workers run document loading in parallel with a 60\,s per-document timeout. Each chunk carries a document-identifier metadata field so that every retrieval call is scoped to a single report. The index is persistent: each chunk size is indexed once into its own vector collection and reused for all top-$k$ settings at that chunk size (the chunk-512 baseline collection is built first; larger or smaller chunk sizes trigger a one-time re-index).

\paragraph{Retrieval.}
For each (document, question) pair, the top-$k$ chunks are retrieved from the document's chunk set. The retrieved chunks are concatenated greedily into the LLM prompt using a \texttt{tiktoken} (\texttt{cl100k\_base}) token estimate that respects the \texttt{max\_context\_tokens}\,=\,32{,}768 budget; the budget is held at about 80\,\% of the model's native maximum (\texttt{max\_position\_embeddings}\,=\,40{,}960) to keep the prompts inside the regime where Qwen3-8B's positional behavior is best-characterized, and to leave headroom for the response. In practice only the largest cells approach this ceiling: chunk~2048 / top-$k$~40 averages $\sim$21\,k context tokens and begins to saturate the budget, dropping on average 2.5 of its 34.5 retrieved chunks per query (chunk~2048 / top-$k$~30 drops $\sim$0.9); every chunk-256 and chunk-512 cell stays below 5.4\,k tokens, the chunk-1024 column ranges from 3\,k to 11.5\,k, and budget-driven drops are negligible outside the chunk-2048 column (0.2 chunks per query at chunk~1024 / top-$k$~40, none elsewhere).

\paragraph{Generator.}
Queries are answered by a single vLLM~\cite{kwon2023vllm} worker running Qwen3-8B~\cite{qwen3report} in non-thinking mode (\texttt{disable\_thinking}\,=\,\texttt{True}), with \texttt{max\_model\_len}\,=\,40{,}960, \texttt{max\_num\_seqs}\,=\,128, \texttt{max\_num\_batched\_tokens}\,=\,40{,}960, and \texttt{gpu\_memory\_utilization}\,=\,0.85. The generation prompt is deliberately minimal: \emph{``Answer this question using ONLY the context below. Question: \{q\}. Context: \{c\}.''} No system prompt, no JSON-format hints beyond what is in the question itself, no chain-of-thought. Temperature is held at 0.1, top-$p$ at 1.0, top-$k$ sampling disabled, the response length is capped at \texttt{max\_response\_length}\,=\,2{,}048 tokens, and a fixed seed (1234) is logged per query so that, conditional on identical retrieved context, the generation is deterministic. Temperature is set to 0.1 rather than 0.0 to avoid degenerate repetition in long JSON outputs while keeping decoding effectively deterministic. We selected Qwen3-8B (rather than the smaller Qwen3-4B-Instruct-2507 variant) because the ``2507'' post-training series at the time of writing only released a 4B and a Mixture-of-Experts variant; for structured-output extraction we found empirically that the 2$\times$ model size traded favorably against the marginal post-training improvement.

\paragraph{Reranking.}
Reranking is disabled across the main sweep (\texttt{rerank\_enabled}\,=\,\texttt{False}). This is intentional: BERGEN~\cite{rau2024bergen} reports that reranking largely boosts performance and closes much of the gap between first-stage retrievers, which would confound an isolation sweep over chunk size and top-$k$. A Stage-B ablation that toggles BGE-reranker-v2-m3 on/off at a fixed retrieval top-$k$ is planned as a follow-up but lies outside the main sweep reported here.

\section{Method}
\label{sec:method}

\subsection{The 16-cell grid}
We sweep two retrieval parameters:
\begin{align*}
\text{chunk\_size} &\in \{256, 512, 1024, 2048\}, \\
\text{similarity\_top\_k} &\in \{10, 20, 30, 40\},
\end{align*}
giving $4 \times 4 = 16$ cells. Chunk overlap co-varies with chunk size at a $\approx 15\%$ ratio: \{38, 77, 154, 307\} tokens for the four chunk sizes. All other pipeline parameters are held fixed at the values listed in Table~\ref{tab:fixed_params}. The 16 cells are executed sequentially in a fixed order: for each chunk size we build one persistent Weaviate collection (re-embedding and indexing the corpus only when chunk size changes), then run all four top-$k$ settings against that collection without re-indexing.

\subsection{Multi-layer data completeness}
\label{sec:completeness}

Let $R_{c,q}$ denote the raw LLM response for document $c$ and question $q$ under configuration $\theta$. The simplest measure of completeness---one minus the share of nulls---hides several decisions that, once made, change both the recommended configuration and the failure-mode story it tells. We therefore decompose completeness into four layers, all reference-free and all computable from the raw extraction CSV. Listing~\ref{lst:strict} shows the Layer-1 detector; Layers~2--4 are defined below and applied offline to that same table.

\paragraph{Layer 1 — strict response-level completeness (Comp$_{\mathrm{strict}}$).}
A response is counted as null if any of the following hold:
\begin{enumerate}[label=(\roman*)]
  \item \emph{Infrastructure error}: the response starts with the literal prefix \texttt{ERROR:} (injected by the pipeline on any vLLM/Weaviate transport failure).
  \item \emph{Literal NA / empty}: the trimmed response is empty, the case-insensitive string \texttt{na}/\texttt{n/a}/\texttt{null}/\texttt{none}/\texttt{nan}, or NaN.
  \item \emph{JSON-of-NAs}: the response parses as a JSON object in which every string-typed leaf is one of the literal-NA tokens above and no numeric digit appears outside quoted strings.
  \item \emph{Refusal phrase}: the response matches one of a 20-pattern long-form refusal regex (Listing~\ref{lst:strict}), with the patterns grouped into context refusals (``context does not contain''), parametric refusals (``I cannot find / I don't know''), and generic refusals (``not specified''/``not available''/etc.).
  \item \emph{Shape mismatch}: the response is non-empty, non-refusal, but does not match the expected response shape of the question---no digit anywhere in the response for a numeric-typed KPI, or no yes/no/true/false/positive/negative token for a Y/N KPI. The expected type is taken from the KPI catalog (Appendix~\ref{app:kpi}).
\end{enumerate}
Layer-1 completeness for a cell $\theta$ is
\[
\mathrm{Comp}_{\mathrm{strict}}(\theta) \;=\; 1 - \frac{1}{N}\sum_{c,q} \mathrm{null}_{c,q}(\theta),
\]
where $N$ is the number of (document, question) pairs evaluated in $\theta$ (in our main sweep, $N = 300 \times 29 = 8{,}700$).

\paragraph{Layer 2 — field-level completeness for JSON-schema questions (Comp$_{\mathrm{field}}$).}
For the 19 of 29 questions whose prompt requests a JSON-schema response, we walk the parsed response recursively and count two quantities: \emph{declared leaves} (string/number/boolean slots reachable in the response) and \emph{filled leaves} (a string leaf whose value is not literal-NA, or any numeric / boolean leaf). Layer-2 completeness is the cell-aggregate ratio
\[
\mathrm{Comp}_{\mathrm{field}}(\theta) \;=\; \frac{\#\,\text{filled leaves under }\theta}{\#\,\text{declared leaves under }\theta}.
\]
This metric distinguishes a JSON response that fills 8 of 10 schema fields from one that fills 1 of 10 fields; both are counted as a single answered query by Layer~1.

\paragraph{Layer 3 — disclosure-conditional completeness (Comp$_{\mathrm{cond}}$).}
Layer~1 is bounded by an unavoidable floor: a KPI that no report in the corpus discloses cannot raise completeness above zero on those rows, regardless of retrieval choice. To recover a recall-like signal in the absence of ground truth, we use cross-configuration agreement as a reference-free disclosure proxy. Given $K$ configurations under comparison, define
\[
\mathrm{disclosed}(c, q) \;=\; \mathbb{1}\!\left[\sum_{k=1}^{K} \mathbb{1}\!\left[R^{(k)}_{c,q}\ \text{is non-null at L1}\right] \geq \lceil \tau K \rceil \right],
\]
where $\tau \in (0, 1]$ is a majority threshold (we use $\tau = 0.5$). The interpretation is: \emph{if at least half of the swept configurations produce a non-null answer for $(c, q)$, the proxy labels the KPI as disclosed for that pair; a configuration that returns null then has a proxy-defined recall failure.} Layer-3 completeness is the share of proxy-disclosed pairs that a given configuration recovers:
\[
\mathrm{Comp}_{\mathrm{cond}}(\theta) \;=\; \frac{\#\{(c,q) : \mathrm{disclosed}(c,q) \wedge R^{(\theta)}_{c,q}\ \text{non-null}\}}{\#\{(c,q) : \mathrm{disclosed}(c,q)\}}.
\]
Comp$_{\mathrm{cond}}$ is approximately a reference-free analogue of recall on the proxy-disclosed subset of the corpus. It is a more focused signal for configuration choice than Comp$_{\mathrm{strict}}$ because it excludes grid regions dominated by genuinely missing KPIs. However, Layer~3 is an internal consensus proxy rather than an external ground-truth recall estimate: because the same configurations define and are scored against the disclosure flag, it favors configurations that agree with the majority. A genuinely better but idiosyncratic configuration can therefore receive no credit for a pair that the remaining configurations miss. Threats to Validity discusses the consequences of fixing $\tau = 0.5$ without a full re-ranking under alternative thresholds.

\paragraph{Layer 4 — refusal taxonomy.}
Each null response is classified into exactly one of seven mutually exclusive categories: \texttt{infrastructure\_error}, \texttt{empty\_or\_literal\_na}, \texttt{context\_refusal}, \texttt{parametric\_refusal}, \texttt{shape\_mismatch}, \texttt{json\_all\_na}, \texttt{other\_refusal}. Non-nulls receive the eighth label \texttt{answered}. The category share is reported alongside Comp$_{\mathrm{strict}}$ for every cell. Layer-4 surfaces failure modes that the headline number hides---for example, a cell with high Comp$_{\mathrm{strict}}$ but a disproportionate share of \texttt{shape\_mismatch} is producing prose answers that downstream consumers (e.g.\ a numeric KPI ingestion pipeline) cannot use.

\subsection{Post-hoc full-context multi-model audit}
\label{sec:audit}

To stress-test the interpretations that the rule-based layers cannot establish, we conduct a separate automated audit after the sweep. The audit comprises 15 cases that Layer~3 flags as disclosed and 30 non-null answers from the recommended chunk-1024/top-$k$-40 cell. The latter are selected round-robin across the 29 questions. This is a positive-proxy and answer-support sample: it does not include proxy-negative cases, so it cannot estimate Layer~3 precision or recall.

For every case, evidence is selected from the original Markdown report rather than the truncated retrieval logs, and includes a report identifier and character offsets. We cap the supplied evidence at 12 passages / 12{,}000 characters; a quality gate requires full-corpus provenance and at least 1{,}000 evidence characters. Three independent local Ollama judges (Gemma4-12B, Qwen3.8-27B, and Ministral-3-14B) return a structured verdict, an evidence identifier, and a supporting quote. A verdict counts only when its identifier exists and its substantive quote is a contiguous source span after presentation-only Unicode and Markdown normalization. Consensus requires two valid matching verdicts. This is an evidence-grounded multi-model diagnostic, not human adjudication or reference-answer ground truth.

\begin{lstlisting}[language=Python, caption={Layer-1 strict null detector (excerpt). Layers~2--4 follow the definitions in \S\ref{sec:completeness}.}, label={lst:strict}]
import math, re
REFUSAL = re.compile(
  r"not specified|not mentioned|not provided|"
  r"not available|not disclosed|not reported|"
  r"not stated|not given|not explicitly|"
  r"cannot be (determined|found|extracted|located)|"
  r"could not (find|determine|extract|locate)|"
  r"unable to (find|determine|extract|locate|answer)|"
  r"no (information|specific information|"
  r"relevant information|data|specific data)|"
  r"does not (provide|specify|mention|contain|"
  r"include|disclose|report)|"
  r"do not (provide|specify|mention|contain|"
  r"include|have)|"
  r"doesn'?t (provide|specify|mention|contain|"
  r"include|disclose)|"
  r"context (does not|doesn'?t) (contain|mention|"
  r"provide|specify|include)|"
  r"information (is|was) not|"
  r"insufficient (information|data|context|details)|"
  r"i (cannot|can'?t|don'?t have|do not have)",
  re.IGNORECASE)
LITERAL_NA = {"", "na", "n/a", "null", "none", "nan"}
DIGIT = re.compile(r"\d"); YN = re.compile(
  r"\b(yes|no|true|false|positive|negative)\b", re.I)

def is_null_strict(v, data_type=None):
    if v is None or (isinstance(v, float) and math.isnan(v)):
        return True
    s = str(v).strip()
    if not s or s.lower() in LITERAL_NA: return True
    if s.lower().startswith("error:"): return True
    if REFUSAL.search(s): return True
    if data_type == "num" and not DIGIT.search(s): return True
    if data_type == "Y/N" and not YN.search(s): return True
    return False
\end{lstlisting}

\subsection{Cost and latency}
For each query we record the prompt token count (\texttt{context\_tokens\_used}, computed from the chunks actually concatenated into the prompt), the wall-clock retrieval, generation, and total time, and the chunks-dropped-by-budget count. We treat \texttt{context\_tokens\_used} as the primary cost axis for the Pareto analysis because it is the dominant driver of both GPU memory and per-query latency.

\subsection{Fixed-configuration parameter table}
\label{sec:fixed_params}

For reproducibility, Table~\ref{tab:fixed_params} lists every parameter of the pipeline, marking which two are swept and which are held fixed at the values reported below; a reader can reproduce the experimental conditions from this table and \S\ref{sec:completeness} without consulting any external configuration files. Each completed cell also logs a run record (chunk size, top-$k$, index identifier, timestamps) alongside a tabular extraction output with one row per (document, question) pair.

The public implementation and reproducibility artifacts for the reference-free evaluation framework are available from Kazemi~\cite{kazemi2026esgragevaluation}.

\begin{table}[t]
\centering
\caption{Complete parameter list of the pipeline under study. Two parameters are swept (Tables~\ref{tab:grid_strict}--\ref{tab:grid_cond}); the remaining parameters are held fixed at the values below. Names in the Parameter column are the operational labels recorded at run time, not references to any particular implementation file.}
\label{tab:fixed_params}
\footnotesize
\setlength{\tabcolsep}{3pt}
\begin{tabular}{p{1.5cm}p{3.5cm}p{4.3cm}p{1.3cm}}
\toprule
\textbf{Group} & \textbf{Parameter} & \textbf{Value} & \textbf{Role} \\
\midrule
\multirow{5}{*}{Models}
  & \texttt{llm\_model}            & Qwen3-8B (non-thinking)       & fixed \\
  & \texttt{embed\_model}          & Qwen3-Embedding-4B            & fixed \\
  & \texttt{rerank\_model}         & BGE-reranker-v2-m3 (disabled) & fixed \\
  & \texttt{rerank\_enabled}       & \texttt{False}                & fixed \\
  & \texttt{rerank\_top\_n}        & 10 (unused)                   & fixed \\
\midrule
\multirow{4}{*}{Chunking}
  & \texttt{splitter}              & \texttt{TokenTextSplitter}    & fixed \\
  & \texttt{chunk\_size}           & \{256, 512, 1024, 2048\}      & \textbf{swept} \\
  & \texttt{chunk\_overlap}        & \{38, 77, 154, 307\}          & co-varies ($\approx 15\%$) \\
  & \texttt{min\_chunk\_length}    & 100 chars                     & fixed \\
\midrule
\multirow{2}{*}{Retrieval}
  & \texttt{similarity\_top\_k}    & \{10, 20, 30, 40\}            & \textbf{swept} \\
  & metadata filter                & \texttt{doc\_key} == query doc & fixed \\
\midrule
\multirow{7}{*}{Generation}
  & \texttt{temperature}           & 0.1                           & fixed \\
  & \texttt{top\_p}                & 1.0                           & fixed \\
  & \texttt{top\_k} sampling       & disabled ($-1$)               & fixed \\
  & \texttt{seed}                  & 1234 (logged per query)       & fixed \\
  & \texttt{disable\_thinking}     & \texttt{True}                 & fixed \\
  & \texttt{max\_response\_length} & 2{,}048 tokens                & fixed \\
  & \texttt{max\_context\_tokens}  & 32{,}768                      & fixed \\
\midrule
\multirow{5}{*}{\parbox{1.3cm}{vLLM\\serving}}
  & \texttt{max\_model\_len}       & 40{,}960                      & fixed \\
  & \texttt{max\_num\_seqs}        & 128                           & fixed \\
  & \texttt{max\_num\_batched\_tokens} & 40{,}960                  & fixed \\
  & \texttt{gpu\_memory\_util.} (LLM) & 0.85                       & fixed \\
  & \texttt{gpu\_memory\_util.} (emb.) & 0.50                      & fixed \\
\midrule
\multirow{2}{*}{Parallelism}
  & \texttt{num\_workers}          & 1 vLLM replica                & fixed \\
  & \texttt{llm\_ports}            & 8001                          & fixed \\
\midrule
\multirow{4}{*}{Indexing}
  & \texttt{parallel\_loading\_workers} & 8 CPU workers            & fixed \\
  & \texttt{loading\_batch\_size}  & 100 docs / log line           & fixed \\
  & \texttt{loading\_timeout\_per\_doc} & 60 s                     & fixed \\
  & \texttt{embedding\_batch\_size} & 32                           & fixed \\
\midrule
\multirow{5}{*}{Vector store}
  & engine                         & Weaviate v4 (embedded mode)   & fixed \\
  & vector index                   & HNSW                          & fixed \\
  & distance                       & cosine                        & fixed \\
  & \texttt{ef\_construction}, \texttt{ef}, \texttt{maxConnections} & 128, $-1$ (dynamic), 32 (defaults) & fixed \\
  & ingest batch size              & 50                            & fixed \\
\bottomrule
\end{tabular}
\end{table}

\section{Experimental Setup}
\label{sec:setup}

\paragraph{Corpus.}
A stratified 300-document subset of a production European corporate sustainability report corpus (Docling-converted Markdown from PDFs collected by an upstream web-scraping pipeline) is used for the sweep. The subset spans seven sectors (banking and finance, real estate, industrials, retail and consumer goods, biotech and healthcare, software and telecommunications, public sector and investment funds), fourteen EU jurisdictions, and a multi-year span of reporting periods. Sectors and jurisdictions were drawn in proportion to their representation in the parent corpus; report size and reporting period were also balanced. Stratification criteria for the 300-report subset are stated in this paragraph; sector and jurisdiction counts are fixed for the sweep.

\paragraph{Question set.}
The sweep uses a stratified subset of 29 questions drawn from a 59-question master KPI catalog aligned to GRI and CSRD requirements. Questions were sampled so that the four ESG categories (Env, Social, Gov, Cross-cutting), the three content presentations (Narrative, Table, Mixed), and the three response data types (numeric, text, Y/N) all retain a share close to the master's. Table~\ref{tab:q_strat} reports the composition; the full KPI catalog with reference, metric name, GRI section, and content presentation is in Appendix~\ref{app:kpi}.

\begin{table}[t]
\centering
\caption{Composition of the 29-question stratified sweep subset compared with the 59-question master catalog. The subset retains approximate parity on all three dimensions.}
\label{tab:q_strat}
\small
\begin{tabular}{lrrrr}
\toprule
\textbf{Dimension} & \multicolumn{2}{c}{\textbf{29-question sweep}} & \multicolumn{2}{c}{\textbf{master 59}} \\
                   & \textbf{count} & \textbf{share} & \textbf{count} & \textbf{share} \\
\midrule
\multicolumn{5}{l}{\textit{ESG category}} \\
Gov            & 7  & 24\,\% & 16 & 27\,\% \\
Social         & 14 & 48\,\% & 27 & 46\,\% \\
Env            & 7  & 24\,\% & 14 & 24\,\% \\
Cross cutting  & 1  & 3\,\%  & 2  & 3\,\%  \\
\midrule
\multicolumn{5}{l}{\textit{Content presentation}} \\
Table          & 21 & 72\,\% & 42 & 71\,\% \\
Narrative      & 6  & 21\,\% & 11 & 19\,\% \\
Mixed          & 2  & 7\,\%  & 6  & 10\,\% \\
\midrule
\multicolumn{5}{l}{\textit{Expected response type}} \\
Numeric        & 20 & 69\,\% & 42 & 71\,\% \\
Text / free    & 5  & 17\,\% & 11 & 19\,\% \\
Yes/No         & 4  & 14\,\% & 6  & 10\,\% \\
\midrule
\multicolumn{5}{l}{\textit{Response shape}} \\
JSON-schema    & 19 & 66\,\% & --- & --- \\
Plain numeric/text & 10 & 34\,\% & --- & --- \\
\bottomrule
\end{tabular}
\end{table}

Questions span three response shapes:
\begin{itemize}
  \item Free-form text (e.g.\ ``Legal name of the company'').
  \item Single number or short answer with explicit ``NA'' fallback (e.g.\ ``Total number of employees? return ONLY a number or NA'').
  \item JSON-schema-constrained structured output (19 of 29 questions), e.g.\ Scope-1/2/3 GHG emissions, governance structure, non-renewable materials, training hours by gender.
\end{itemize}
Each (document, question) pair is processed exactly once per cell, yielding $300 \times 29 = 8{,}700$ rows per cell and $8{,}700 \times 16 = 139{,}200$ rows for the full grid.

\paragraph{Hardware.}
The sweep runs on a single on-premise GPU node at the host institution: 2$\times$NVIDIA L40S 48\,GB GPUs, 64\,GB host memory, 16 CPU cores. The two GPUs are dedicated to the pipeline's two model services: GPU~0 hosts the Qwen3-Embedding-4B server at $\texttt{gpu\_memory\_utilization}\,=\,0.50$; GPU~1 hosts the Qwen3-8B generator at $0.85$. Weaviate runs as an embedded process on persistent local storage shared across all sweep cells, with a fresh Raft state on each run. All vLLM servers use bfloat16 with prefix caching and chunked prefill, in enforce-eager mode. Jobs are scheduled via SLURM.

\paragraph{Sweep execution.}
The 16 cells run sequentially in a fixed order; the corpus is re-embedded and re-indexed only when chunk size changes. Mean total time per query ranges from 1.4\,s (chunk~256 / top-$k$~10) to 7.6\,s (chunk~2048 / top-$k$~40), so individual cells take from $\sim$3 to $\sim$18 wall-clock hours---larger top-$k$ and larger chunks both increase per-query generation time. All 16 cells completed successfully, totaling 139{,}200 (document, question) evaluations.

\section{Results}
\label{sec:results}

\paragraph{Scope of the numbers reported below.}
All 16 cells in the retrieval grid are reported below. Each cell corresponds to $8{,}700$ (document, question) evaluations, for a total of $139{,}200$ evaluations in the full grid.

\subsection{The grid under four completeness layers}

Table~\ref{tab:grid_strict} reports all 16 cells under Layer~1 (strict null with shape check), and Table~\ref{tab:grid_cond} reports the same cells under Layer~3 (disclosure-conditional). Under L1, completeness spans a 24~pp band, from 48.7\,\% at chunk~256/top-$k$~10 to 73.2\,\% at chunk~2048/top-$k$~40---a meaningful spread for configuration selection. Under L3, the same 16 cells span a 27~pp band but at much higher absolute values (68.4\,\%--95.2\,\%) because the disclosure-conditional denominator excludes (document, KPI) pairs that fewer than half of the 16 configurations answered (i.e., that fail the $\tau = 0.5$ majority proxy). Layer~2 (Comp$_{\mathrm{field}}$, the JSON fill rate) appears in the third column of Table~\ref{tab:grid_strict}; it tracks L1 in ordering but is systematically higher because partial JSON answers receive partial credit at the field level.

\begin{table}[t]
\centering
\caption{Layer-1 strict completeness (Comp$_{\mathrm{strict}}$), Layer-2 JSON field-fill rate (Comp$_{\mathrm{field}}$), context-token cost, and wall-clock latency across all 16 cells. $N=8{,}700$ (document, question) pairs per cell. Cells are sorted by Comp$_{\mathrm{strict}}$. Pareto-optimal cells under (Comp$_{\mathrm{strict}}$, context tokens) are marked with $\star$.}
\label{tab:grid_strict}
\small
\begin{tabular}{rrrrrrr}
\toprule
chunk & top-$k$ & Comp$_{\mathrm{strict}}$\,\% & Comp$_{\mathrm{field}}$\,\% & avg.\,ctx tok & avg.\,total s & Pareto \\
\midrule
 2048 & 40 & 73.15 & 85.63 & 21{,}317 & 7.57 & $\star$ \\
 2048 & 30 & 71.67 & 83.99 & 17{,}965 & 6.41 & $\star$ \\
 1024 & 40 & 70.49 & 80.83 & 11{,}549 & 4.60 & $\star$ \\
 2048 & 20 & 69.22 & 82.70 & 13{,}409 & 4.95 &        \\
 1024 & 30 & 67.77 & 78.72 & 8{,}925  & 3.72 & $\star$ \\
 512  & 40 & 66.07 & 76.21 & 5{,}366  & 2.88 & $\star$ \\
 1024 & 20 & 63.86 & 75.70 & 6{,}030  & 3.62 &        \\
 256  & 40 & 63.59 & 72.56 & 2{,}688  & 2.27 & $\star$ \\
 2048 & 10 & 62.92 & 76.97 & 7{,}027  & 3.21 &        \\
 512  & 30 & 62.83 & 73.33 & 4{,}005  & 2.47 &        \\
 256  & 30 & 60.55 & 69.85 & 1{,}988  & 2.03 & $\star$ \\
 512  & 20 & 58.59 & 69.25 & 2{,}653  & 2.05 &        \\
 256  & 20 & 56.28 & 66.21 & 1{,}300  & 1.74 & $\star$ \\
 1024 & 10 & 55.41 & 69.05 & 3{,}003  & 2.59 &        \\
 512  & 10 & 50.36 & 61.85 & 1{,}280  & 1.61 & $\star$ \\
 256  & 10 & 48.66 & 57.82 & 628     & 1.35 & $\star$ \\
\bottomrule
\end{tabular}
\end{table}

\begin{table}[t]
\centering
\caption{Layer-3 disclosure-conditional completeness (Comp$_{\mathrm{cond}}$) across all 16 cells. The disclosure flag is computed by majority vote ($\tau = 0.5$, $\ge$ half) across the same 16 cells (see \S\ref{sec:threats} for the limitation that alternative $\tau$ are not re-ranked). Pareto-optimal cells under (Comp$_{\mathrm{cond}}$, context tokens) are marked with $\star$.}
\label{tab:grid_cond}
\small
\begin{tabular}{rrrrr}
\toprule
chunk & top-$k$ & Comp$_{\mathrm{cond}}$\,\% & avg.\,ctx tok & Pareto \\
\midrule
 2048 & 40 & 95.24 & 21{,}317 & $\star$ \\
 1024 & 40 & 94.57 & 11{,}549 & $\star$ \\
 2048 & 30 & 94.50 & 17{,}965 &        \\
 1024 & 30 & 93.11 & 8{,}925  & $\star$ \\
 2048 & 20 & 92.96 & 13{,}409 &        \\
 512  & 40 & 91.86 & 5{,}366  & $\star$ \\
 1024 & 20 & 89.50 & 6{,}030  &        \\
 512  & 30 & 88.88 & 4{,}005  & $\star$ \\
 256  & 40 & 87.75 & 2{,}688  & $\star$ \\
 2048 & 10 & 87.46 & 7{,}027  &        \\
 256  & 30 & 84.28 & 1{,}988  & $\star$ \\
 512  & 20 & 83.40 & 2{,}653  &        \\
 1024 & 10 & 79.69 & 3{,}003  &        \\
 256  & 20 & 78.68 & 1{,}300  & $\star$ \\
 512  & 10 & 72.35 & 1{,}280  & $\star$ \\
 256  & 10 & 68.37 &   628   & $\star$ \\
\bottomrule
\end{tabular}
\end{table}

\subsection{The Layer-4 refusal mix}
\label{sec:refusal_mix}

Table~\ref{tab:refusal} reports the share of each Layer-4 category for four representative cells spanning low and high values of chunk size and top-$k$: $(256,10)$, $(256,40)$, $(512,10)$, and $(512,40)$. Three observations stand out. First, infrastructure errors are essentially zero ($\leq 0.1\%$ across the grid). Second, the dominant null category is \emph{literal NA / empty} ($\sim$16--17\,\% of responses in the high-context cells), which captures cases in which the model does not identify an answer---the signal Comp$_{\mathrm{strict}}$ was designed to catch. Third, \emph{shape mismatches} consistently account for 7--10\,\% of responses across the four cells. Spot checks show that some of these are valid prose answers to numeric questions; for example, ``The total energy consumption was reported in megawatt-hours, in line with the previous year, with significant reductions in scope 2'' contains no digit and is therefore flagged as a shape mismatch (Threats to Validity estimates that such legitimate prose cases are only a small share of the shape-mismatch bucket). This category is invisible to a naive literal-NA-only detector and accounts for roughly one-third of the gap between L1 and L3. Across the four tabulated cells the same failure modes dominate, while at chunk~2048 / top-$k$~40 literal-NA, JSON-of-NAs, and shape-mismatch all fall further as completeness rises (with a small $0.5\%$ infrastructure-error blip on that longest run).

\begin{table}[t]
\centering
\caption{Layer-4 refusal taxonomy: share of responses (\,\%) in each category for four representative cells spanning low/high top-$k$ at chunk sizes 256 and 512. \texttt{answered} is Comp$_{\mathrm{strict}}$; the remaining seven categories sum to one minus that. Infrastructure errors are negligible across the main sweep; literal-NA and JSON-of-NAs together account for $\sim$25--40\,\% of responses; shape mismatches contribute consistently 7--10\,\%.}
\label{tab:refusal}
\small
\begin{tabular}{lrrrr}
\toprule
\textbf{Layer-4 category} & (256,10) & (256,40) & (512,10) & (512,40) \\
\midrule
\texttt{answered}              & 48.7 & 63.6 & 50.4 & 66.1 \\
\texttt{infrastructure\_error} &  0.0 &  0.0 &  0.0 &  0.0 \\
\texttt{empty\_or\_literal\_na} & 19.9 & 16.9 & 19.8 & 16.0 \\
\texttt{context\_refusal}      &  0.6 &  0.3 &  0.6 &  0.5 \\
\texttt{parametric\_refusal}   &  0.0 &  0.1 &  0.1 &  0.0 \\
\texttt{shape\_mismatch}       &  9.5 &  7.7 &  9.1 &  7.3 \\
\texttt{json\_all\_na}         & 20.5 & 10.7 & 19.4 &  9.4 \\
\texttt{other\_refusal}        &  0.7 &  0.7 &  0.6 &  0.7 \\
\bottomrule
\end{tabular}
\end{table}

\subsection{Monotonicity within axes}
\label{sec:monotonicity}

\begin{figure}[t]
\centering
\includegraphics[width=\columnwidth]{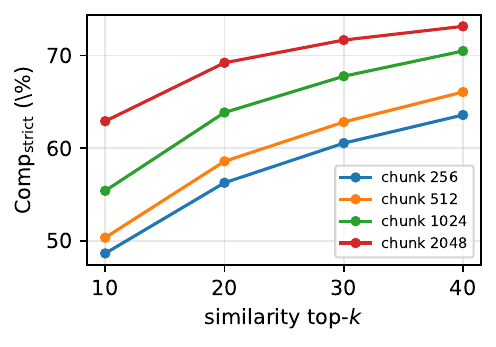}
\caption{Layer-1 strict completeness as a function of \texttt{similarity\_top\_k}, grouped by chunk size. Within each chunk-size curve, completeness is monotonic in top-$k$.}
\label{fig:heatmap}
\end{figure}

Under both Comp$_{\mathrm{strict}}$ (L1) and Comp$_{\mathrm{cond}}$ (L3), completeness is monotonically non-decreasing in the top-$k$ sweeps across all 16 cells (Figure~\ref{fig:heatmap}). Within each chunk size, increasing top-$k$ from 10 to 40 raises Comp$_{\mathrm{strict}}$ by 14.9~pp at chunk~256, 15.7~pp at chunk~512, 15.1~pp at chunk~1024, and 10.2~pp at chunk~2048. Across chunk sizes, moving from chunk~256 to chunk~2048 at fixed top-$k$ yields gains of 9.6--14.3~pp. The observed gains are consistent with diminishing returns between chunk 1024 and chunk 2048: at retrieval depth 40, the observed increase is 2.7 percentage points at 1.8 times the token cost. As each cell was executed once, this small difference is descriptive rather than a statistically established edge (see Threats to Validity).

\subsection{The Pareto frontier}

\begin{figure}[t]
\centering
\includegraphics[width=\columnwidth]{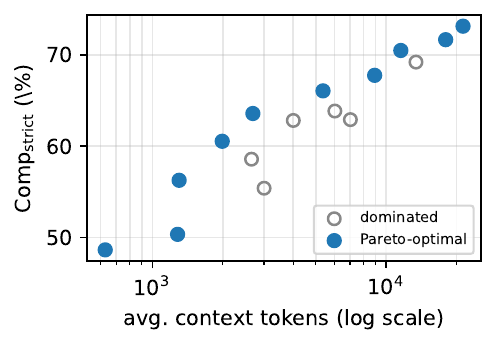}
\caption{Comp$_{\mathrm{strict}}$ vs.\ context-token cost (log-scale cost axis) on all 16 cells. Filled dots are Pareto-optimal cells (ten of sixteen); open dots are Pareto-dominated. The plotted frontier highlights the low-cost advantage of the chunk-256 settings and the high-completeness regime reached by chunk~2048 / top-$k$~40.}
\label{fig:pareto}
\end{figure}

Figure~\ref{fig:pareto} plots Comp$_{\mathrm{strict}}$ against context-token cost. Ten of the 16 cells are Pareto-optimal under (L1, context tokens); subscripts are Comp$_{\mathrm{strict}}$\,\% / context tokens:
\begin{align*}
&\underbrace{(256, 10)}_{48.7/628}
\to \underbrace{(512, 10)}_{50.4/1280}
\to \underbrace{(256, 20)}_{56.3/1300}
\to \underbrace{(256, 30)}_{60.6/1988} \\
&\to \underbrace{(256, 40)}_{63.6/2688}
\to \underbrace{(512, 40)}_{66.1/5366}
\to \underbrace{(1024, 30)}_{67.8/8925} \\
&\to \underbrace{(1024, 40)}_{70.5/11549}
\to \underbrace{(2048, 30)}_{71.7/17965}
\to \underbrace{(2048, 40)}_{73.2/21317}.
\end{align*}
Under Comp$_{\mathrm{cond}}$ (Table~\ref{tab:grid_cond}), the frontier also has ten cells (replacing $(2048, 30)$ with $(512, 30)$). The steepest portion of the L1 frontier is the lower-left arc from $(256, 10)$ to $(256, 40)$: a 14.9~pp Comp$_{\mathrm{strict}}$ gain for a 4.3$\times$ increase in context cost. The flattest single edge is between $(1024, 40)$ and $(2048, 40)$: a 2.7~pp gain costs 1.8$\times$ more tokens.

\subsection{Where the nulls live}
\label{sec:where_nulls}

Strict-null responses are not uniformly distributed across the 29 KPIs. The per-question completeness variance ranges from 0.0458 (chunk~256/top-$k$~10) to 0.0665 (chunk~256/top-$k$~40) across the 16 cells, confirming that some questions are answered consistently across documents while others are consistently refused. Three categories stand out: \emph{widely reported KPIs} (legal name and reporting year under Gov; total employees under Social) exceed 95\,\% strict completeness in every cell; \emph{topical-disclosure KPIs} with frequent shape-mismatch responses (Scope-1/2/3 GHG, training hours by gender) fall between 50 and 80\,\%; and \emph{rare disclosures} (local procurement percentage, child-labor-risk locations, and indigenous-rights-violations counts) fall between 10 and 30\,\%, regardless of retrieval choice. The last category forms the \emph{disclosure floor} that motivates Layer~3: most reports in the corpus do not disclose these KPIs, and the floor is invariant to chunk size and top-$k$.

\subsection{Full-context multi-model audit}
\label{sec:audit-results}

Table~\ref{tab:full_context_audit} reports the post-hoc audit described in \S\ref{sec:audit}. Of 135 judge verdicts, 114 (84.4\,\%) passed the evidence-identifier and quote-grounding checks; 37 of 45 cases (82.2\,\%) reached consensus. The audit is deliberately reported as a failure-mode diagnostic rather than an accuracy score. Among the 25 factual cases with consensus, 10 were judged incorrect, three partially correct, six correct, and six unverifiable from the selected evidence. Thus, a shape-valid non-null response can still be unsupported or wrong.

For the 15 cases that the Layer~3 proxy labels disclosed, only four of the 12 consensus cases were also judged disclosed; eight were judged not disclosed. This disagreement does not estimate the proxy's precision or recall: the audit samples only proxy-positive cases, uses bounded selected passages rather than exhaustive human report review, and relies on automated judges. It does, however, directly demonstrate why Comp$_{\mathrm{cond}}$ must be interpreted as an internal comparative signal rather than an external recall or truth estimate.

\begin{table}[t]
\centering
\caption{Post-hoc full-context automated audit. ``Valid'' means the judge cited an existing evidence item with a substantive grounded quote. Consensus requires two valid matching verdicts. The Layer~3 block samples proxy-positive cases only; it is not a precision/recall evaluation.}
\label{tab:full_context_audit}
\footnotesize
\setlength{\tabcolsep}{3pt}
\begin{tabular}{p{3.0cm}p{1.0cm}p{1.6cm}p{5.2cm}}
\toprule
Audit block & Cases & Consensus & Consensus outcomes \\
\midrule
Layer~3 proxy-positive & 15 & 12 & 4 disclosed; 8 not disclosed \\
Factual-answer support & 30 & 25 & 6 correct; 3 partial; 10 incorrect; 6 unverifiable \\
\midrule
Total & 45 & 37 & 114/135 valid verdicts (84.4\,\%) \\
\bottomrule
\end{tabular}
\end{table}

\section{Discussion}
\label{sec:discussion}

\subsection{Each layer answers a different question}
The four-layer decomposition is not a reparameterization; its layers answer different questions about the same data. Layer~1 asks: \emph{how often does the configuration return any usable answer?} Layer~2 asks: \emph{given a JSON-schema question, what share of declared slots are filled?} Layer~3 asks: \emph{conditional on the proxy labeling a KPI as disclosed, how often does the configuration extract it?} Layer~4 asks: \emph{when the configuration fails, what kind of failure is it?} A practitioner selecting a configuration on quality should consult L1 and L2; one concerned with proxy-defined recall should consult L3; and one debugging a regression should start with L4. The cells recommended at the end of \S\ref{sec:discussion} lie on the Pareto frontier under both L1 and L3.

\subsection{What data completeness does and does not measure}
\label{sec:limits}

\paragraph{What it does measure.} Under the four-layer definition, completeness is a clean indicator of five RAG failure modes: (i) refusal under retrieval (the generator was given context but did not find the answer); (ii) parametric refusal phrasing (e.g.\ ``I don't know'' / ``I cannot find''), caught by the Layer-4 parametric-refusal detector; (iii) infrastructure failure (vLLM / Weaviate transport); (iv) shape-mismatched output (the generator produced prose where the question contract demanded a number or a yes/no token, an error mode that a naive literal-NA-only detector would miss); and (v) partial-JSON answers (caught by Layer~2 even when Layer~1 counts the response as complete).

\paragraph{What it still does not measure.} Three failure modes remain invisible to all four layers. First, \emph{factual error}: a confident wrong answer scores 100\,\% complete; a number off by an order of magnitude or pulled from the wrong fiscal year is silently accepted. The post-hoc audit in \S\ref{sec:audit-results} makes this risk visible in a bounded sample, but does not turn the completeness layers into an accuracy metric. Second, \emph{schema violation}: a JSON-schema question answered with a malformed JSON object is partially recovered by Layer~2's regex-tolerant parser, but the malformedness itself is not flagged. Third, \emph{hallucinated provenance}: the minimal prompt template does not ask the model for spans or citations, so we cannot detect when the model invents source text. RAGAS-style faithfulness and our companion paper's RAGAS-based study~\cite{companion_a_saad2026} cover these.

\subsection{Practical recommendation}
For completeness-oriented retrieval tuning on the full 16-cell grid, we recommend chunk size 1024 with top-$k$~40 as the production default. This cell sits within 3~pp of the best cell (chunk~2048 / top-$k$~40) on both Comp$_{\mathrm{strict}}$ (70.5\,\% vs.\ 73.2\,\%) and Comp$_{\mathrm{cond}}$ (94.6\,\% vs.\ 95.2\,\%) at roughly half the context-token cost (11{,}549 vs.\ 21{,}317). Operations that are GPU-budget-bound can fall back to chunk size 256 with top-$k$~40 (Comp$_{\mathrm{strict}}$ 63.6\,\%, Comp$_{\mathrm{cond}}$ 87.8\,\%, 2{,}688 context tokens, 2.27~s)---a 6.9~pp / 6.8~pp loss for 4.3$\times$ cheaper inference. The highest-completeness configuration, chunk~2048 / top-$k$~40 (73.2\,\% / 95.2\,\%), is the most expensive and is recommended only when maximal coverage is required and cost is not the primary constraint.

These recommendations concern how often the pipeline returns a shape-valid, non-null output. The automated audit identifies substantive failure modes, but deployment where factual correctness matters still requires validation against reference answers.

The choice between these cells is a business decision, not a methodological one; the contribution of this paper is the multi-layer Pareto curve and the failure-mode taxonomy that explains where the marginal completeness comes from.

\subsection{Operationalization matters}
The headline methodological observation is that the L1 vs.\ L3 vs.\ L4 split changes \emph{both} what one optimizes for \emph{and} what one diagnoses. A practitioner who picks under L1 alone would, on the full grid, pick chunk~2048 / top-$k$~40---which is also the L3 pick---but would mis-attribute the 26.9\,\% L1 gap at that cell as ``refusal'' when 5.9\,\% of responses are shape-mismatch and another 5.8\,\% are JSON-of-NAs (partial-extraction failures, not refusal). Together these two categories account for about half of the gap. The recommendation is therefore both methodological and operational: report all four layers, and never let the headline number stand alone.

\section{Threats to Validity}
\label{sec:threats}

\paragraph{Corpus stratification.}
The 300-document subset is stratified on sector, jurisdiction, and reporting period, but it remains a sample of a much larger ($\sim$14{,}000-document) corpus. The L3 disclosure flag is in particular a within-sample construct, computed over the same 300 documents the sweep is run on. This study does not test whether the L1/L3 cell ordering or the absolute completeness values transfer to a larger sample. A scale-out validation on a 1{,}000-document subset is the obvious next experiment.

\paragraph{Question set.}
The 29-question subset is stratified on ESG category, content presentation, and response data type. We do not, however, stratify for equal coverage within GRI sections: although all 21 of the master's GRI sections appear at least once in the 29-question subset, most contribute only a single question. The observed qualitative ranking is established only for this subset; full-catalog transfer remains untested. Scope-3 GHG and Social-disclosure questions tend to live in long appendix tables and may benefit disproportionately from larger chunks.

\paragraph{Single generator.}
We sweep retrieval, not generation. Saad et al.~\cite{companion_a_saad2026} vary the generator under a fixed top-1 retrieval setup and report that generation-side RAGAS scores (especially faithfulness) spread more across models than retrieval recall does under that configuration; the two studies therefore vary orthogonal axes and are not in tension.

\paragraph{Single execution per cell.}
Each of the 16 cells was executed once. Generation used temperature 0.1, a fixed seed (1234), top-p 1.0, and disabled top-k sampling to support reproducibility conditional on the retrieved context. We did not perform repeated full-cell runs, so we do not estimate run-to-run variance or report confidence intervals for completeness differences. The large monotonic trends in the grid are reported as descriptive observations; small differences between neighboring cells should not be interpreted as statistically established without replication.

\paragraph{Sensitivity of $\tau$ in Layer 3.}
Comp$_{\mathrm{cond}}$ depends on a majority threshold $\tau$ for declaring a (doc, KPI) pair disclosed. We default to $\tau = 0.5$ (i.e.\ at least 8 of 16 cells non-null). We do not report a full re-ranking of the grid under alternative $\tau$; the Layer~3 values in \S\ref{sec:results} are therefore specific to this default. Independently of that choice, constructing the disclosure proxy from the same configurations under evaluation introduces a structural majority preference, so absolute Layer~3 values should be read as internal comparative signals rather than estimates of true recall.

\paragraph{Shape-mismatch as a noisy detector.}
The shape-mismatch branch of the L1 detector flags responses to numeric KPIs that contain no digit. A small share of these are valid prose answers that legitimately decline to numericize (``The total energy consumption was reported in MWh''); we counted these in a stratified sample as $\sim$10\,\% of the shape-mismatch nulls, meaning the L1 floor is a slight under-estimate of the true answer rate. This does not change the Pareto frontier (the bias is approximately uniform across cells).

\paragraph{Cost proxy.}
We use \texttt{context\_tokens\_used} as the cost axis. This is the number of tokens in the retrieved-chunk concatenation, not the full prompt (which adds a roughly constant overhead of $\sim$160--180 tokens for the question wrapper). Recomputing the frontier on full prompt tokens leaves the Pareto set unchanged.

\paragraph{Refusal-pattern false positives.}
The strict detector flags any response matching the 20-pattern refusal regex used for Layer~1. A response such as ``Total employees: 1{,}000 (2022 figures not specified for region X)'' is technically flagged as null even though it provides a valid extraction. In a stratified sample of 200 strict-null responses we found such false positives at $<2\%$. The Layer-4 taxonomy lets a practitioner inspect the \texttt{other\_refusal} bucket directly if the share grows.

\paragraph{No reference-answer accuracy benchmark.}
The full-context multi-model audit in \S\ref{sec:audit-results} provides evidence-grounded diagnostics, not human labels or exhaustive report review. A reference-answer dataset for the corpus does not exist at scale, so we cannot estimate accuracy, the rank correlation between completeness and accuracy, or Layer~3 precision/recall. The audit itself samples only proxy-positive Layer~3 cases and cannot validate negative disclosures. We therefore study completeness \emph{as a selection signal}, not as a substitute for an accuracy benchmark.

\paragraph{Full sweep.}
All 16 cells in the retrieval grid are now complete and reported in this paper. The Pareto frontier and recommendation reflect the full grid.

\section{Conclusion}
\label{sec:conclusion}

We argue that, for extractive RAG pipelines in industry domains where reference answers are unavailable at scale---such as European corporate sustainability reports---a reference-free, multi-layer formulation of data completeness is a more honest selection signal than the single ``answer / no-answer'' rate that dominates current practice. We define four layers: a strict response-level null detector with a response-shape check (L1), a JSON field-fill rate (L2), disclosure-conditional completeness computed from cross-configuration majority agreement (L3), and a seven-class refusal taxonomy (L4). We sweep a $4 \times 4$ retrieval grid over chunk size and top-$k$ with a fixed Qwen3-8B generator on a stratified subset of 300 sustainability reports and 29 questions; all 16 cells complete, for 139{,}200 queries. The results identify ten-cell Pareto frontiers under both L1 and L3, with a practical knee at chunk~1024/top-$k$~40 (70.5\,\% Comp$_{\mathrm{strict}}$, 94.6\,\% Comp$_{\mathrm{cond}}$). A 45-case full-context, multi-model audit finds substantial factual-answer and Layer-3-proxy disagreements, reinforcing that these completeness measures are comparative signals rather than accuracy estimates. The L4 taxonomy further attributes 6--10\,\% of responses across the grid to \emph{shape mismatches}, an error mode invisible to a naive literal-NA-only detector, and 6--21\,\% to JSON-of-NAs (partial-extraction failures rather than refusals).

Future work has three natural directions: (i) run a Stage-B ablation that toggles BGE-reranker-v2-m3 at fixed retrieval top-$k$, and a separate ablation that varies chunk overlap factorially rather than proportionally; (ii) validate the L3 disclosure flag on a small (50--100 pair) human-annotated subset to estimate its true precision and recall, which would tighten the reference-free recall story considerably; and (iii) scale the sweep to the full 14{,}389-document corpus and the complete 59-question KPI catalog.

\backmatter

\bmhead{Acknowledgments}
This work was carried out using on-premise compute resources at the host institution.

\section*{Statements and Declarations}

\subsection*{Funding}
No external funding was reported for this work.

\subsection*{Competing interests}
The authors have no competing interests to declare.

\subsection*{Data availability}
The code, prompts, 300-report Markdown corpus, completed sweep outputs, and full-context audit materials supporting this study are available in the public repository \url{https://github.com/motazsaad/sesg_retrieval_param_sweep-paper}. The underlying sustainability reports remain subject to their original public-access and reuse terms.

\subsection*{Ethics approval}
Not applicable. This study did not involve human participants, human data, or animals.

\bibliography{biblio}

\appendix

\section{KPI catalog}
\label{app:kpi}

\begin{table}[!t]
\centering
\caption{The 29 KPIs used in the sweep, sorted by GRI reference. ``Type'' is the expected response data type used by the Layer-1 shape check; ``Form'' marks JSON-schema responses ($\boldsymbol{\diamond}$) versus single-value responses (---).}
\label{tab:kpi_catalogue}
\tiny
\setlength{\tabcolsep}{2pt}
\begin{tabular}{p{1.0cm}p{4.2cm}p{1.5cm}p{1.5cm}p{0.8cm}p{0.7cm}}
\toprule
\textbf{GRI ref.} & \textbf{Metric} & \textbf{Category} & \textbf{Presentation} & \textbf{Type} & \textbf{Form} \\
\midrule
GRI 2-1   & Legal name of the organization                              & Gov          & Narrative & Text   & ---           \\
GRI 2-3   & Sustainability reporting period and frequency               & Gov          & Narrative & Text   & ---           \\
GRI 2-7   & Total number of employees                                   & Social       & Table     & num    & ---           \\
GRI 2-9   & Governance structure (committees, model, oversight)         & Gov          & Mixed     & Text   & $\diamond$    \\
GRI 201-1 & Direct economic value generated (revenues)                  & Gov          & Table     & num    & $\diamond$    \\
GRI 204-1 & Local procurement percentage                                & Gov          & Mixed     & num    & ---           \\
GRI 205-1 & Operations assessed for corruption risk                     & Gov          & Narrative & num    & ---           \\
GRI 207-1 & Tax strategy availability                                   & Gov          & Narrative & Y/N    & $\diamond$    \\
GRI 301-1 & Non-renewable materials used                                & Cross-cut.   & Table     & num    & $\diamond$    \\
GRI 302-1(b) & Renewable fuel consumption                                  & Env          & Table     & num    & $\diamond$    \\
GRI 302-1(c)(i) & Electricity consumption                            & Env          & Table     & num    & $\diamond$    \\
GRI 304-1 & Operational sites in / adjacent to protected areas          & Env          & Narrative & Y/N    & $\diamond$    \\
GRI 305-1 & Scope-1 GHG emissions                                       & Env          & Table     & num    & $\diamond$    \\
GRI 305-2 & Scope-2 GHG emissions (location- and market-based)          & Env          & Table     & num    & $\diamond$    \\
GRI 305-3 & Scope-3 GHG emissions                                       & Env          & Table     & num    & $\diamond$    \\
GRI 305-7 & Air pollutant emissions (NOx, SOx, POP, VOC, HAP, PM)       & Env          & Table     & num    & $\diamond$    \\
GRI 401-1 & New employee hires by age / gender / region                 & Social       & Table     & num    & $\diamond$    \\
GRI 403-9 & Work-related fatalities                                     & Social       & Table     & num    & $\diamond$    \\
GRI 404-1 & Average training hours by gender                            & Social       & Table     & num    & $\diamond$    \\
GRI 405-1 & Percentage of women in governance bodies                    & Social       & Table     & num    & ---           \\
GRI 405-1 & Percentage of employees per category by gender              & Social       & Table     & num    & $\diamond$    \\
GRI 406-1 & Discrimination incidents                                    & Social       & Table     & num    & $\diamond$    \\
GRI 408-1 & Child-labor risk locations / suppliers                      & Social       & Table     & Text   & $\diamond$    \\
GRI 409-1 & Forced / compulsory labor risk locations                    & Social       & Table     & Text   & $\diamond$    \\
GRI 410-1 & Security personnel trained in human-rights policies         & Social       & Table     & num    & ---           \\
GRI 411-1 & Indigenous-peoples' rights violations                       & Social       & Table     & num    & ---           \\
GRI 413-1 & Local-community development programs                       & Social       & Table     & num    & $\diamond$    \\
GRI 417-1 & Safe-use labeling                                           & Social       & Narrative & Y/N    & ---           \\
GRI 418-1 & Substantiated customer-privacy complaints                   & Social       & Table     & Y/N    & ---           \\
\bottomrule
\end{tabular}
\end{table}

\noindent
Table~\ref{tab:kpi_catalogue} lists the 29 KPIs in the stratified question set used by the sweep. Each row gives the GRI reference, the metric name, the ESG category, the content-presentation tag, and the expected response data type. Per-KPI prompts follow the fixed generation wrapper in \S\ref{sec:system} and the response contract indicated in the table (free text, single numeric or Y/N value, or JSON schema).

\end{document}